\documentclass[12pt,twoside,english,a4paper,prd,nofootinbib,preprint,showpacs,preprintnumbers,floatfix]{revtex4}
\usepackage{pslatex}
\usepackage[T1]{fontenc}
\usepackage{geometry}
\usepackage{graphicx}
\usepackage{amssymb}

\makeatletter

\newcommand{\lyxdot}{.}

\usepackage{babel}
\makeatother
\begin{document}
\begin{flushright}PITHA 08/29\\
MZ-TH/08-38\par\end{flushright}

\vspace{4mm}

\begin{center}\textbf{\LARGE Determining the CP parity of Higgs bosons}\par\end{center}{\LARGE \par}

\begin{center}\textbf{\LARGE at the LHC}\par\end{center}{\LARGE \par}

\begin{center}\textbf{\LARGE in the $\tau$ to 1-prong decay channels}\par\end{center}{\LARGE \par}

\begin{center}\vspace{6mm}
\par\end{center}

\begin{center}\textbf{\large Stefan Berge}$^{*}$\textbf{\large }%
\footnote{\texttt{\small berge@uni-mainz.de}%
} \textbf{\large and \, Werner Bernreuther}$^{\dagger}$\textbf{\large }%
\footnote{\texttt{\small breuther@physik.rwth-aachen.de}%
} \textbf{\large }\par\end{center}{\large \par}

\begin{center}$^{*}$Institut für Physik (WA THEP), Johannes Gutenberg-Universität,
55099 Mainz, Germany \par\end{center}

\begin{center}\vspace{-15mm}
\par\end{center}

\begin{center}$^{\dagger}$ Institut für Theoretische Physik, RWTH
Aachen University, 52056 Aachen, Germany \par\end{center}

\begin{center}\vspace{17mm}
 \textbf{Abstract}\par\end{center}

We propose a method for determining the $CP$ nature of a neutral
Higgs boson or spin-zero resonance $\phi$ at the CERN Large Hadron
Collider (LHC) in its $\phi\to\tau^{-}\tau^{+}$ decay channel. The
method can be applied to any 1-prong $\tau$-decay mode, which comprise
the majority of the $\tau$-lepton decays. The proposed observables
allow to discriminate between pure scalar and pseudoscalar Higgs-boson
states and/or between a $CP$-conserving and $CP$-violating Higgs
sector. We show for the decays $\tau\to\pi\nu_{\tau}$ that the method
maintains its discriminating power when measurement uncertainties
are taken into account. The method will be applicable also at a future
linear $e^{+}e^{-}$ collider.

\vspace{35mm}

PACS numbers: 11.30.Er, 12.60.Fr, 14.80.Bn, 14.80.Cp\\
 Keywords: hadron collider physics, Higgs bosons, tau leptons, parity,
CP violation

\section{\textbf{Introduction} }

The major physics goal at the CERN Large Hadron Collider (LHC) is
the search for Higgs bosons or other (spin-zero) resonances that pin
down the mechanism of electroweak gauge symmetry breaking. (For reviews,
see, e.g., \cite{Djouadi:2005gi,Djouadi:2005gj,GomezBock:2007hp,Hill:2002ap}.)
If such particles are found, then the next task would be the exploration
of their properties. For electrically neutral spin-zero states this
includes the determination of the parity $(P)$ and charge conjugation
times parity $(CP)$ quantum numbers, respectively, which provide
important information about the dynamics of these particles. There
is an extensive literature on proposals of how to measure the $CP$
properties of Higgs bosons in their production and decay processes
at hadron colliders or at a future linear $e^{+}e^{-}$ collider,
including \cite{Dell'Aquila:1985ve,Dell'Aquila:1988rx,Dell'Aquila:1988fe,Bernreuther:1993df,Bernreuther:1993hq,Bernreuther:1997af,Bernreuther:1998qv,Chang:1993jy,Barger:1993wt,Kramer:1993jn,Grzadkowski:1995rx,Klamke:2007cu,Berge:2008wi,Desch:2003rw,Rouge:2005iy,Plehn:2001nj,Buszello:2002uu,BhupalDev:2007is}.
(For a recent compilation and overview, see \cite{Accomando:2006ga}.)

A promising reaction in this respect is Higgs-boson decay into $\tau$
lepton pairs, where $\tau$-spin correlations discriminate between
$CP$-even and -odd Higgs-boson states, and between a $CP$-conserving
and $CP$-violating Higgs sector. For subsequent $\tau$-decays into
three charged prongs it was shown in \cite{Berge:2008wi} that
experimentally robust discriminating observables exist also for the LHC. 
 In order to substantially
increase the data sample in future experiments, one would like to
employ for the measurement of the $CP$ properties of a Higgs boson
also $\tau$ decays into one charged prong. However, for these modes
the method proposed in \cite{Berge:2008wi} is not applicable at the
LHC, because it requires the reconstruction of the $\tau^{\mp}$ rest
frames. In this letter we construct observables that can also be applied
to 1-prong $\tau$ decays. We demonstrate by simulations taking expected
measurement uncertainties into account that, at the LHC, the $CP$
nature of a neutral Higgs boson -- or any neutral spin-zero boson
which decays into $\tau^{-}\tau^{+}$ pairs -- can be determined with
these observables.

\section{Observables\label{sec-obs}}

The analysis below applies to any neutral spin-zero resonance $h_{j}$,
in particular to any neutral Higgs boson, with flavor-diagonal couplings
to quarks and leptons $f$ (with mass $m_{f}$) 
\begin{equation}
{\cal
  L}_{Y}=-(\sqrt{2}G_{F})^{1/2}\sum_{j,f}m_{f}(a_{jf}\bar{f}f+b_{jf}\bar{f}i\gamma_{5}f)\, h_{j} \,\, ,
\label{bbzyukl}
\end{equation}
 where $G_{F}$ is the Fermi constant and $a_{jf}$ and $b_{jf}$
are the model-dependent reduced scalar and pseudoscalar Yukawa couplings.
In the SM, $a_{f}=1$ and $b_{f}=0$. SM extensions where the couplings 
 (\ref{bbzyukl}) appear include  models with two Higgs
doublets, such as the non-supersymmetric type II models and the minimal
supersymmetric SM extension (MSSM) (see, e.g., \cite{Djouadi:2005gi,Djouadi:2005gj,Accomando:2006ga,GomezBock:2007hp}).
These models contain three physical neutral Higgs fields $h_{j}$
in the mass basis. If Higgs sector $CP$ violation (CPV) is negligibly
small, then the fields $h_{j}$ describe two scalar states $h,H$
($b_{jf}=0)$ and a pseudoscalar $A$ $(a_{jf}=0)$. In the case of
Higgs sector CPV, the $h_{j}$ are $CP$ mixtures, that is, they have
non-zero couplings $a_{jf}$ and $b_{jf}$ to quarks and leptons (see,
e.g., \cite{Bernreuther:1992dz}) which lead to $CP$-violating effects
in $h_{j}\to f{\bar{f}}$ already at the Born level.

In the following, $\phi$ denotes any of the neutral Higgs bosons
$h_{j}$ just discussed or, more generally, a neutral spin-zero resonance.
The observables discussed below for determining the $CP$ quantum
number of $\phi$ in its $\tau$-decay channel may be applied to any
Higgs production process $i\to\phi+X\to\tau^{-}+\tau^{+}+X.$ At the
LHC, this includes the gluon and gauge boson fusion processes $gg\to\phi$
and $q_{i}q_{j}\to\phi\, q'_{i}q'_{j}$, respectively, and the associated
production $t{\bar{t}}\phi$ or $b{\bar{b}}\phi$ of a light Higgs
boson $\phi$. The spin of the resonance can be determined in standard
fashion from the polar angle distribution of the $\tau$ leptons.
In order to determine with these reactions whether $\phi$ is a scalar,
a pseudoscalar, or a $CP$ mixture, one can use $\tau^{\mp}$ spin
correlations. They lead to specific angular correlations among the
charged particles (charged prongs) from $\tau^{-}$ and $\tau^{+}$
decay. A suitable set of observables involves the opening angle distribution
between the charged prongs, $CP$-odd triple correlations and asymmetries
\cite{Bernreuther:1993df,Bernreuther:1997af}. In order to exploit
the full discriminating power of these observables, one must be able
to determine the $\tau^{\mp}$ rest frames, i.e., the energies and
three-momenta of the $\tau$ leptons. At the LHC, the reconstruction
of the $\tau^{\mp}$ rest frames is possible for $\tau$ decays into
3 charged prongs, $\tau^{-}\to2\pi^{-}\pi^{+}\nu_{\tau}$ and likewise
for $\tau^{+}$. For these channels, it was shown in \cite{Berge:2008wi}
that one can discriminate, at the LHC, with these observables i) between
a scalar, a pseudoscalar $\phi$, and a $CP$ mixture and also ii)
between (nearly) mass-degenerate scalar and pseudoscalar Higgs bosons
with $CP$-invariant couplings and one or several $CP$ mixtures.

In order to increase the statistics, one would like to exploit also
$\tau$ decays into one charged particle. We consider here the case
where both $\tau^{-}$ and $\tau^{+}$ decay into one charged prong.
Then the determination of the $\tau^{\mp}$ four-momenta is not possible
without further assumptions. However, for our purpose, it is not necessary
to reconstruct the $\tau$ rest frames. As we will show below, one
can construct discriminating observables in the zero-momentum frame
(ZMF) of the two charged prongs from the $\tau^{\mp}$ decays which
involve only directly measurable quantities, namely the momenta of
the charged prongs and the impact parameter vectors defined below.
For definiteness, we consider in the following the case where both
$\tau^{-}$ and $\tau^{+}$ decay into a charged pion and a neutrino,
\begin{equation}
p\, p\to\phi+X\to\tau^{-}\tau^{+}+X\to\pi^{-}\pi^{+}+X \,.
\label{LHCphipi}
\end{equation}
In order to put our approach into perspective, we first briefly recapitulate
another method for determining the $CP$ nature of $\phi$ in this
$\tau$ decay channel. It was pointed out a long time ago \cite{Dell'Aquila:1988rx}
that the distribution of the angle between the normal vectors of the
$\tau^{-}$ and $\tau^{+}$ decay planes discriminates between a $CP=+1$
and $CP=-1$ Higgs boson. The formula of \cite{Dell'Aquila:1988rx}
can be generalized to the case where $\phi$ has arbitrary scalar
and pseudoscalar couplings to $\tau$ leptons. We consider, in the
$\phi$ rest frame, the decay 
\begin{equation}
\phi\to\tau^{-}({\bf k}^{\phi})\,+\,\tau^{+}({\bf
  -k}^{\phi})\to\pi^{-}({\bf p}_{-}^{\tau})\,+\,\pi^{+}({\bf
  p}_{+}^{\bar{\tau}})\,+\,\nu_{\tau}+\bar{\nu}_{\tau} \, .
\label{phidectau}
\end{equation}
 Here ${\bf k}^{\phi}$ is the 3-momentum of the $\tau^{-}$ in the
rest frame of $\phi$, and ${\bf p}_{-}^{\tau}\,({\bf p}_{+}^{\bar{\tau}})$
is the $\pi^{-}\,(\pi^{+})$ 3-momentum in the $\tau^{-}\,(\tau^{+})$
rest frame. We shall take the $\tau^{-}$ direction ${\bf \hat{k}}^{\phi}$
as $z$ axis both in the $\tau^{-}$ and the $\tau^{+}$ rest frame.
Denoting the azimuthal angle of the $\pi^{-}\,(\pi^{+})$ in the $\tau^{-}\,(\tau^{+})$
rest frame by $\varphi_{-}\,(\varphi_{+})$, one notices that $\varphi=\varphi_{+}-\varphi_{-}$
is the angle between the normal vectors of the $\tau^{-}\to\pi^{-}$
and $\tau^{+}\to\pi^{+}$ decay planes spanned by the above momentum
vectors. Using the $\phi\to\tau^{-}\tau^{+}$ spin-density matrix
\cite{Bernreuther:1997af} and the SM density matrix of polarized
$\tau\to\pi\nu$ decay, we obtain: 
\begin{equation}
\Gamma^{-1}\frac{d\Gamma}{d\varphi}=\frac{1}{2\pi}[1-\frac{\pi^{2}}{16}(c_{1}\cos\varphi+c_{2}\sin\varphi)]\,,
\label{decplcorr}
\end{equation}
 where $0\leq\varphi<2\pi$ and
\begin{equation}
c_{1}=\frac{a_{\tau}^{2}\beta_{\tau}^{2}-b_{\tau}^{2}}{a_{\tau}^{2}\beta_{\tau}^{2}+b_{\tau}^{2}}\,,\qquad
c_{2}=-\frac{2a_{\tau}b_{\tau}\beta_{\tau}}{a_{\tau}^{2}\beta_{\tau}^{2}+b_{\tau}^{2}}\,.
\label{c1c3}
\end{equation}
(For similar considerations, see \cite{Desch:2003rw,Rouge:2005iy}.)
Here $a_{\tau},\ b_{\tau}$ are the couplings defined in (\ref{bbzyukl}).
In (\ref{c1c3}) the velocity $\beta_{\tau}$ may be put equal to
1. Eq. (\ref{decplcorr}) includes the special cases of a pure scalar
$(c_{1}=1,\, c_{2}=0)$ and of a pure pseudoscalar $(c_{1}=-1,\, c_{2}=0)$,
where the distribution is proportional to $1\mp({\pi^{2}}/{16})\cos\varphi.$
In the case of an ideal $CP$ mixture, $a_{\tau}=\pm b_{\tau}$, the
distribution takes the form $(2\pi)^{-1}(1\pm({\pi^{2}}/{16})\sin\varphi).$
While for a pure scalar or pseudoscalar $\phi$ the complete information
on (\ref{decplcorr}) is contained already in the range $0\leq\varphi<\pi$,
one must determine (\ref{decplcorr}) in the complete 
 interval $0\leq\varphi<2\pi$
in order to check for $CP$ violation \cite{Desch:2003rw}. For this
aim, one must define/measure signed normal decay-plane vectors. If
one cannot distinguish $\varphi$ from $2\pi-\varphi$, the resulting
distribution of the angle $\varphi$ between the unsigned normal vectors
is
\begin{equation}
\Gamma^{-1}\frac{d\Gamma}{d\varphi}=\frac{1}{\pi}(1-\frac{\pi^{2}}{16}c_{1}\cos\varphi)\
,\label{eq:dec1cor}
\end{equation}
where here $0\leq\varphi<\pi.$ (Eq. (\ref{eq:dec1cor}) is obtained
by adding (\ref{decplcorr}) evaluated at $\varphi$ and at $2\pi-\varphi.$)
In (\ref{eq:dec1cor}) the parity-odd term has averaged out. 

At the LHC, it is extremely difficult - if not impossible - to measure
the distributions (\ref{decplcorr}), (\ref{eq:dec1cor}), because
the determination of the $\phi$ and $\tau^{\mp}$ rest frames requires
the reconstruction of the $\tau$ energies and momenta in the laboratory
frame. Even for the simplest $\tau^- \tau^+$ decay  channel
(\ref{LHCphipi}) the energies of the tau leptons
need still to be fixed, using the missing momentum ${\bf p}_{T}^{miss}$
in the plane transverse to the proton beam. This leads to large uncertainties.

\begin{figure}[h]
 \begin{centering}\includegraphics[scale=0.5]{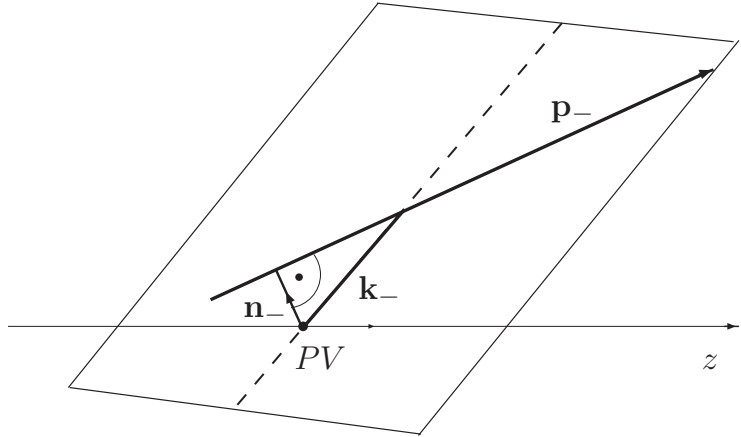} \par
\end{centering}
\caption{Definition of the impact parameter vector ${\bf n}_{-}$ in the plane
of the decay $\phi\to\tau^{-}\to\pi^{-}$ in the laboratory frame.
Here, ${\bf p}_{-}$ is the measured $\pi^{-}$ momentum, $PV$ is
the $\tau^{-}$ production vertex, and ${\bf k}_{-}$ is the 3-momentum
of the $\tau^{-}$. \label{fig:decplL}}
\end{figure}

In order to proceed, we notice that the distributions (\ref{decplcorr})
and (\ref{eq:dec1cor}) remain invariant -- in the absence of detector
cuts -- when we switch from the $\tau^{-}\tau^{+}$ ZMF to another
inertial frame, the $\pi^{-}\pi^{+}$ ZMF. Of course, the determination
of the (un)signed decay-plane correlation in this frame requires knowledge
of the $\tau^{\mp}$ momenta, too. As this is not feasible in general,
we propose to use instead two observables in this frame, which can
be unambiguously determined from quantities measured in the laboratory
frame. The joint use of these observables avoids also the determination
of a signed correlation. We construct these observables in the following
way: \\
 1) Consider the $\tau^{\mp}$ decays in the laboratory frame. For
$\tau^{-}\to\pi^{-}$ the decay plane in this frame is shown in Fig.~\ref{fig:decplL}.
It is determined by the measured $\pi^{-}$ direction of flight and
by the $\tau^{-}\tau^{+}$ production vertex $PV$, which is practically
equal to the Higgs boson production vertex. This vertex is obtained
from the visible tracks of the charged particles/jets produced in
association with the Higgs boson $\phi$ \cite{Gennai:2006}. The
$\tau^{+}\to\pi^{+}$ decay plane in the laboratory frame is obtained
in analogous fashion. One can now determine the impact parameter vectors
${\bf n}_{\mp}$ in the laboratory frame by projecting perpendicularly
onto the $\pi^{\mp}$ directions from $PV$ -- see Fig.~\ref{fig:decplL}.
The pion momenta and the impact parameter vectors fix the normal vectors
of these two decay planes. The distribution of the angle between these
normal vectors or, alternatively, the distribution of the angle $\varphi_{lab}$
between the vectors ${\bf n}_{-}$ and ${\bf n}_{+}$ shows already
some sensitivity for discriminating between a scalar and a pseudoscalar
boson -- see the next section. 
2) A much higher sensitivity can be achieved by determining the analogous
correlations in the $\pi^{-}\pi^{+}$ ZMF. One can reconstruct this
frame by a Lorentz boost from the laboratory frame with the measured
pion 4-momenta $p_{\mp}^{\mu}=(E_{\mp},{\bf p}_{\mp})$. The resulting
$\pi^{\mp}$ energies and momenta are $E_{\mp}^{*},\,{\bf p}_{\mp}^{*}$
with ${\bf p}_{+}^{*}=-{\bf p}_{-}^{*}$. (All quantities in this
frame will be denoted by an asterisk.) However, the true decay planes
in this frame can not be reconstructed, because the true impact parameter
vectors in this frame can not be obtained from the measured laboratory-frame
3-vectors ${\bf n}_{\mp}$. Instead we proceed as follows. Denoting
the normalized impact parameter vectors in the laboratory frame by
${\bf \hat{n}}_{\mp}$, we \textit{define} the two space-like laboratory-frame
4-vectors $n_{\mp}^{\mu}=(0,{\bf \hat{n}}_{\mp})$. These vectors
are boosted to the $\pi^{-}\pi^{+}$ ZMF, and we obtain $n_{\mp}^{*\mu}=(n_{0\mp}^{*},{\bf n}_{\mp}^{*})$.
Next we decompose the spatial parts ${\bf n}_{\mp}^{*}$ into components
parallel and perpendicular to the respective pion momentum ${\bf p}_{\mp}^{*}$:
\begin{equation}
{\bf n}_{\mp}^{*}=r_{\perp}^{\mp}{\bf
  \hat{n}}_{\perp}^{*\mp}+r_{\parallel}^{\mp}{\bf
  \hat{n}}_{\parallel}^{*\mp}\,,
 \label{eq: nperppar}
\end{equation}
 where $r_{\perp}^{\mp},r_{\parallel}^{\mp}$ are constants. In this
way we obtain the unit vectors ${\bf \hat{n}}_{\perp}^{*\mp}$, which
are orthogonal to ${\bf p}_{\mp}^{*}$, respectively, for each event
in a unique fashion. The angle, which takes the role of the true angle
between the unsigned normal vectors of the decay planes , 
Eq. (\ref{eq:dec1cor}),
is defined by 
\begin{equation}
\varphi^{*}=\arccos({\bf \hat{n}}_{\perp}^{*+}\cdot{\bf
  \hat{n}}_{\perp}^{*-})\,,
 \label{eq:apprangle}
\end{equation}
where $0\leq\varphi^{*}<\pi.$ In addition, the $CP$-odd and $T$-odd
triple correlation ${\cal O}_{CP}^{*}={\bf \hat{p}}_{-}^{*}\cdot({\bf \hat{n}}_{\perp}^{*+}\times{\bf \hat{n}}_{\perp}^{*-})$
turns out to be an appropriate tool for distinguishing between $CP$
invariance and $CP$ violation in Higgs-boson decay. Here ${\bf \hat{p}}_{-}^{*}$
denotes the normalized $\pi^{-}$ momentum. As $-1\leq{\cal O}_{CP}^{*}\leq1$,
it is convenient to consider, alternatively, the distribution of the
angle 
\begin{equation}
\psi_{CP}^{*}=\arccos({\bf \hat{p}}_{-}^{*}\cdot({\bf
  \hat{n}}_{\perp}^{*+}\times{\bf
  \hat{n}}_{\perp}^{*-}))\,.
 \label{eq:cpangle}
\end{equation}
 We shall show in the next section that (\ref{eq:apprangle}) and
(\ref{eq:cpangle}) are sensitive and robust observables for determining
the $CP$ nature of a neutral Higgs boson.

\section{Results}

As already emphasized above, the observables (\ref{eq:apprangle})
and (\ref{eq:cpangle}) can be applied to the $\tau$-decay channel
of any Higgs-boson production process. The reason is that the normalized
distributions of these variables do not depend on the Higgs-boson
momentum if no detector cuts are applied. Furthermore we shall show
for $\phi\to\tau^{-}\tau^{+}\to\pi^{-}\pi^{+}$ that detector cuts
have only a  small effect on these distributions for Higgs masses
larger than $200$~GeV. Thus, our results will not change significantly
if one considers a different Higgs production mode or if initial-state
higher-order QCD corrections are taken into account. Therefore, we
have computed in this analysis all distributions for the LHC reaction
(\ref{LHCphipi}) with a Higgs boson production process at leading
order. Specifically we have used $gg\to\phi$ and $b\bar{b}\to\phi$.
For non-standard Higgs bosons $\phi$ and large $\tan\beta$, the
latter production mode, respectively $gg\to b\bar{b}\phi,$ is considered
to be the most promising one in the search for the $\phi\to\tau\bar{\tau}$
decay channel at the LHC \cite{:1999fr,Ball:2007zza}.

\begin{figure}[h]
 \begin{centering}\includegraphics[clip,scale=0.44]{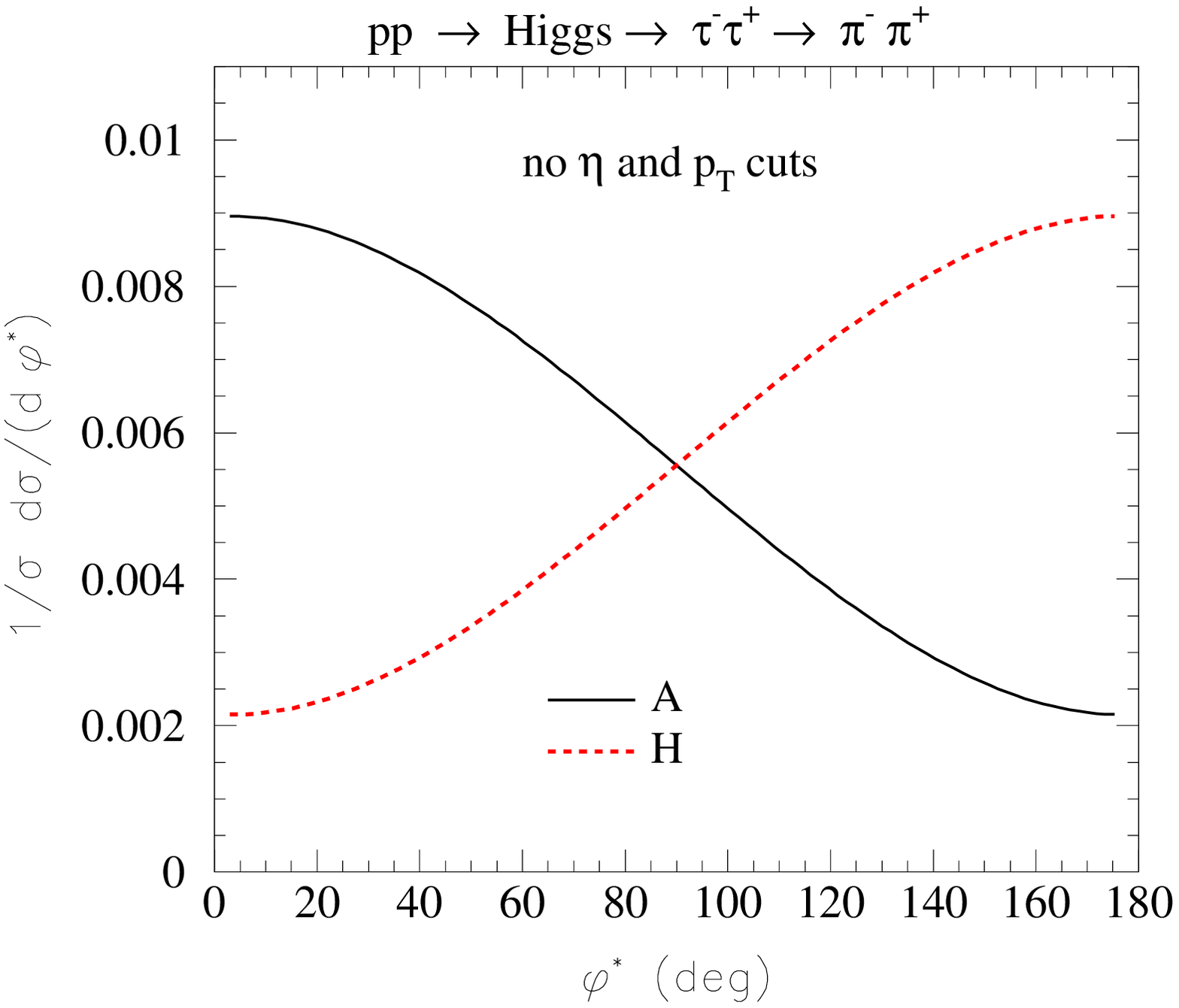}\hspace*{1.cm}\includegraphics[clip,scale=0.44]{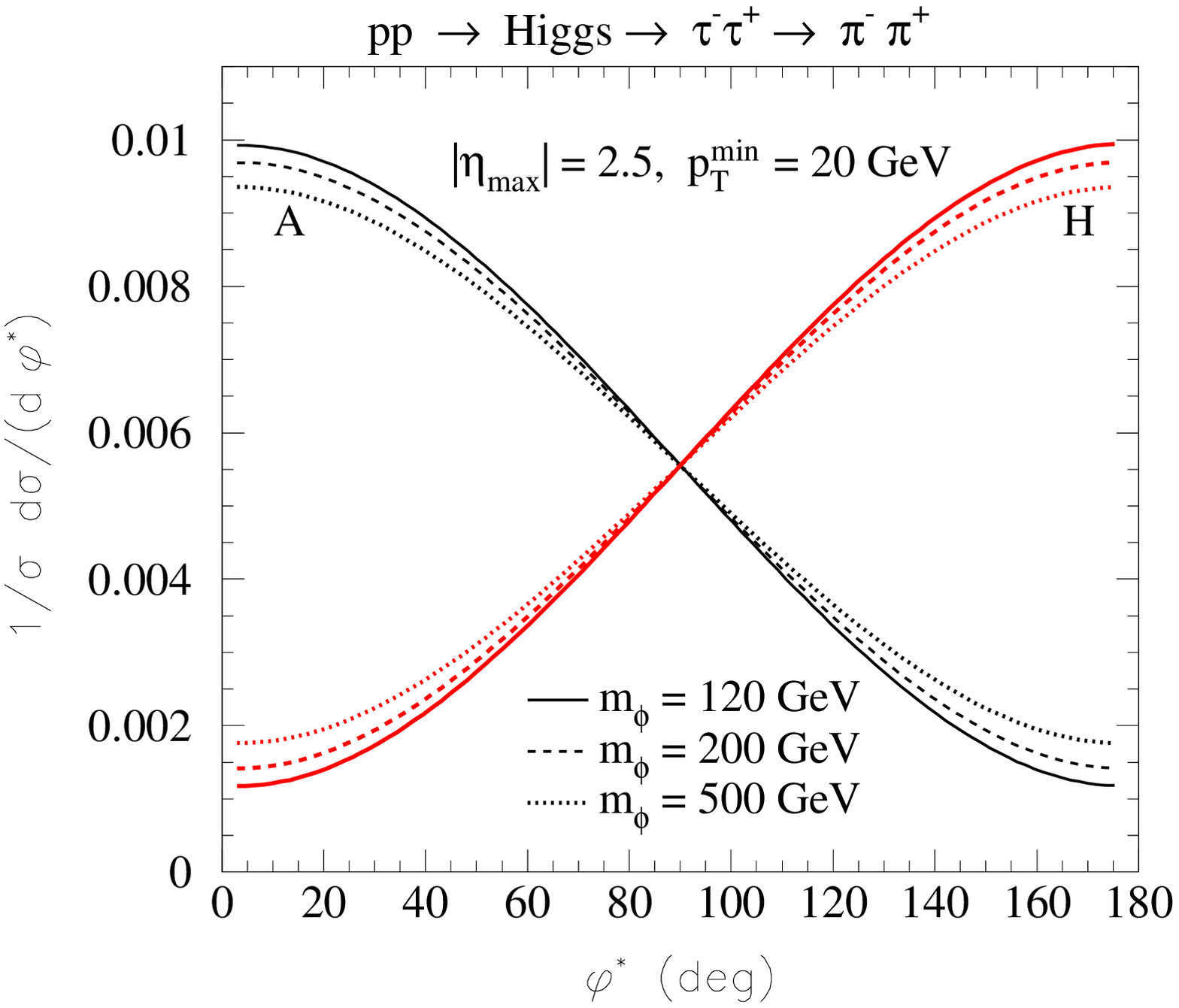}\vspace{-3pt}
 \par
\end{centering}
\caption{(a) The distribution of (\ref{eq:apprangle}) in the $\pi\pi$ ZMF
for a scalar and pseudoscalar Higgs boson without detector cuts. (b)
Dependence of the $\varphi^{*}$ distribution on the Higgs-boson mass
if detector cuts are applied. \label{fig:2ab}}
\end{figure}

Fig.~\ref{fig:2ab} (a) shows the distribution of the angle (\ref{eq:apprangle})
for a scalar $(\phi=H)$ and a pseudoscalar $(\phi=A)$ Higgs boson,
which is determined according to the procedure described above, in
the absence of detector cuts. 
We checked for Higgs-boson masses $120\,\,{\rm GeV}\leq m_{\phi}\leq500\,\,{\rm GeV}$
that this distribution is practically independent of $m_{\phi}$.
Moreover, this distribution is practically identical to the distribution
of the true decay-plane angle $\sigma^{-1}d\sigma/d\varphi_{true}^{*}=(\pi)^{-1}(1\mp(\pi^{2}/16)\cos\varphi_{true}^{*})$
in the $\pi\pi$ ZMF (see (\ref{eq:dec1cor})), which could be determined
if the $\tau^{\mp}$ four-momenta in the laboratory frame were known.

Next we apply cuts on the $\pi^{\mp}$ pseudo-rapidities, $|\eta|\leq2.5$,
and on their transverse momenta, $p_{T}=\sqrt{p_{x}^{2}+p_{y}^{2}}\geq20$
GeV, and recompute this distribution for various Higgs-boson masses.
Fig.~\ref{fig:2ab} (b) shows that it depends only very weakly on
$m_{\phi}$, both for $\phi=H$ and $\phi=A$. 

\begin{figure}[h]
\begin{centering}\includegraphics[clip,scale=0.44]{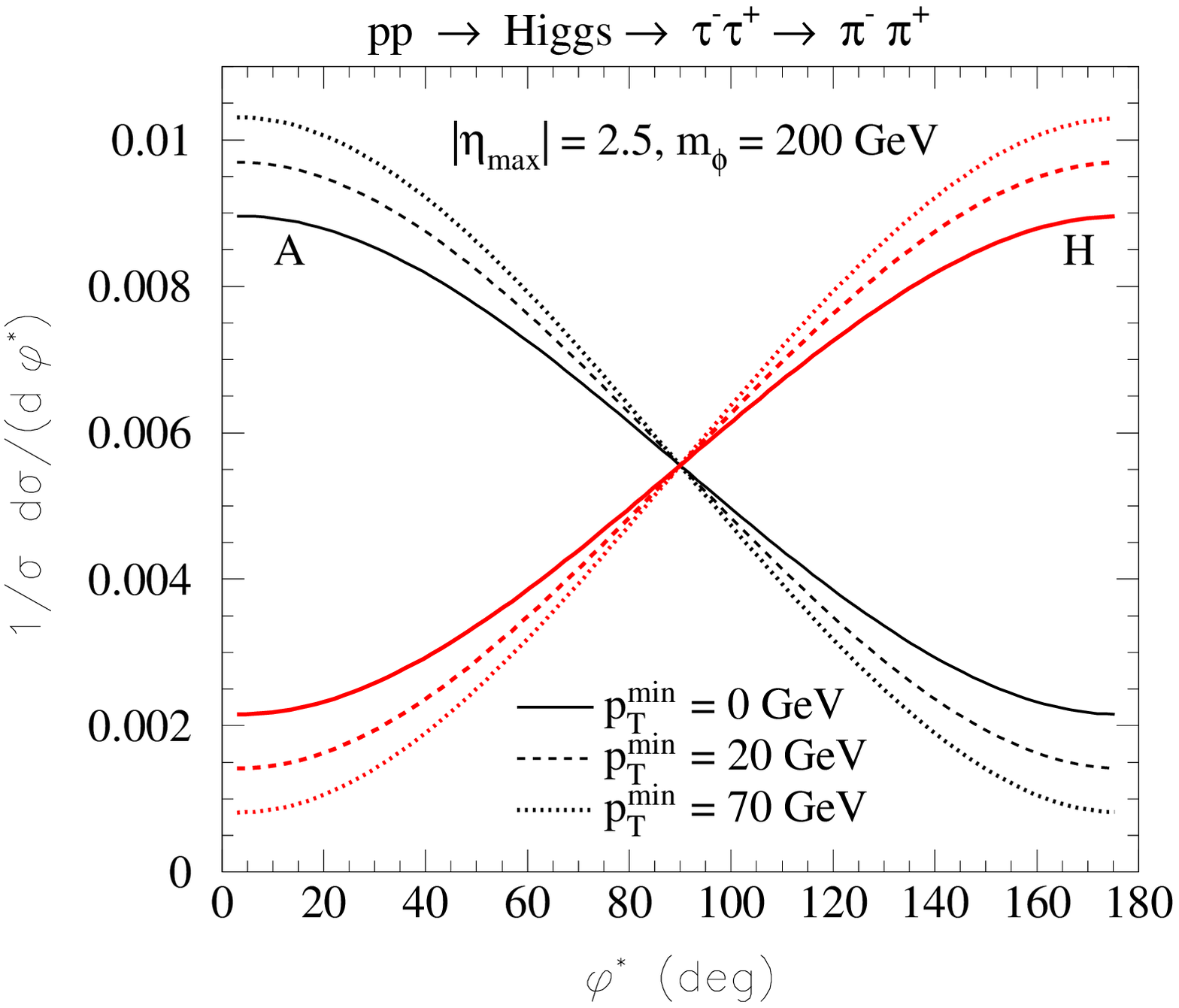}\hspace*{1.cm}\includegraphics[clip,scale=0.44]{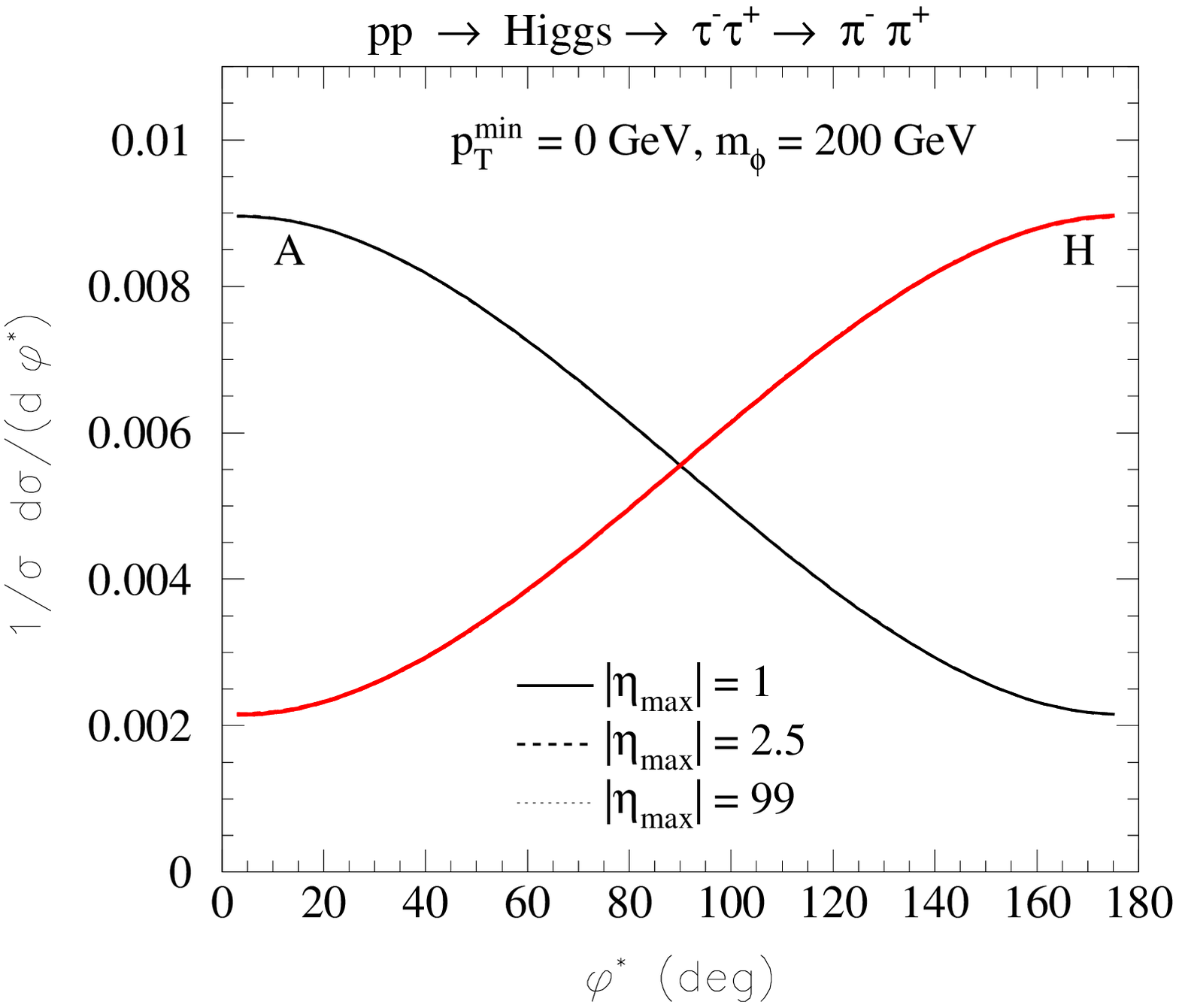}\vspace{-3pt}
\par
\end{centering}
\caption{(a) Dependence of the $\varphi^{*}$-distribution in the $\pi\pi$
ZMF on the required minimal transverse pion momentum (a), and on the
maximal pseudo-rapidity (b). In the latter case, the curves lie on
top of each other. \label{fig:3ab}}
\end{figure}

In Figs.~\ref{fig:3ab} (a), (b) we have plotted, both for $\phi=H$
and $\phi=A$, the dependence of the $\varphi^{*}$-distribution on
the cut on the $\pi^{\mp}$ transverse momenta and on $\eta$, respectively,
for $m_{\phi}=200$~GeV. While there is a relatively weak dependence
on $p_{T}^{min}$, the dependence on $\eta_{max}$ is negligibly small.
We checked for $120\,{\rm GeV}\leq m_{\phi}\leq500\,{\rm GeV}$ that
this feature holds true also for other Higgs-boson masses.

Performing studies for the distribution 
$\sigma^{-1}d\sigma/d\varphi_{true}^{*}$
of the true decay-plane angle $\varphi_{true}^{*}$ analogous
 to Fig.~\ref{fig:2ab}
(b) and Figs.~\ref{fig:3ab} (a), (b) we find that it is practically
identical to $\sigma^{-1}d\sigma/d\varphi^{*}$ also in these cases.
This demonstrates that the angle $\varphi^{*}$ is a very efficient
variable for discriminating between a $CP$-even and $CP$-odd Higgs-boson
in a wide range of masses $m_{\phi}$.

\begin{figure}[h]
 \begin{centering}\includegraphics[clip,scale=0.42]{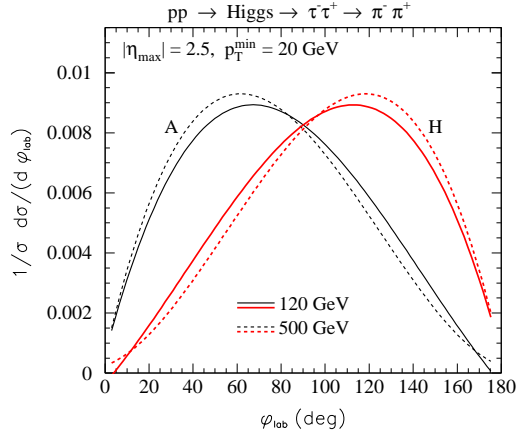}
\vspace{-3pt}
\par
\end{centering}
\caption{The distribution of the angle $\varphi_{lab}$ in the laboratory
frame for a scalar and a pseudoscalar Higgs boson with mass $m_{H,A}=120$
GeV and $m_{H,A}=500$ GeV. \label{fig:4}}
\end{figure}

One may wonder whether the angle 
$\varphi_{lab}=\arccos({\bf n}_{+}\cdot{\bf n}_{-})$
between the impact-parameter vectors ${\bf n}_{+}$ and ${\bf n}_{-}$
in the laboratory frame is already sensitive to the Higgs-boson parity.
In Fig. \ref{fig:4} the distribution of this angle is plotted for
a scalar and a pseudoscalar boson with mass $m_{H,A}=120$ GeV and
$500$ GeV, respectively. Fig. \ref{fig:4} shows that $\varphi_{lab}$
has some sensitivity: in the case of $H$ decay the events
peak around $\varphi_{lab}=120^{\circ}$, while for $A$ decay the
maximum of the distribution is near $\varphi_{lab}=60^{\circ}$. It
is gratifying that the dependence of the distributions on the mass
of the Higgs boson is not very strong. For light states the distance
between the maxima of the $A$ and $H$ distributions is somewhat
smaller than in the case of heavy states. This is due to the fact
that lighter Higgs bosons have, for a specific production reaction,
a larger average velocity in the laboratory frame than heavy states.
The larger the speed of the Higgs boson, the more the discriminating
power of the $\varphi_{lab}$ distribution will be diminished.
A comparison of Fig. \ref{fig:4} with Figs. \ref{fig:2ab}, \ref{fig:3ab}
shows that the distribution of the angle $\varphi^{*}$in the $\pi\pi$
ZMF is, nevertheless, more sensitive to the parity of a Higgs boson
than the distribution of $\varphi_{lab}.$ Therefore we continue to
analyze the former. 

\begin{figure}[h]
 \begin{centering}\includegraphics[clip,scale=0.42]{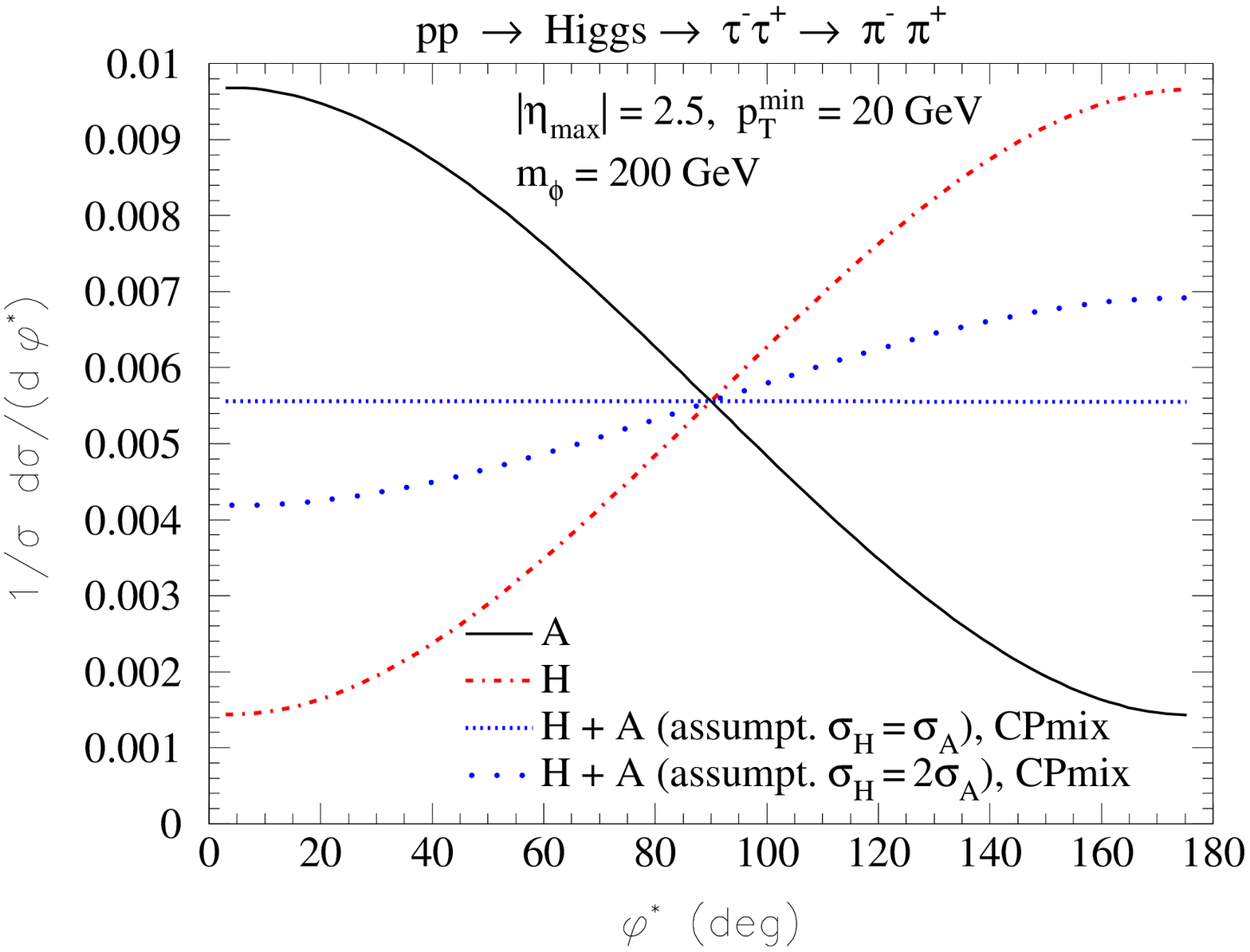}\hspace*{.4cm}
\includegraphics[clip,scale=0.42]{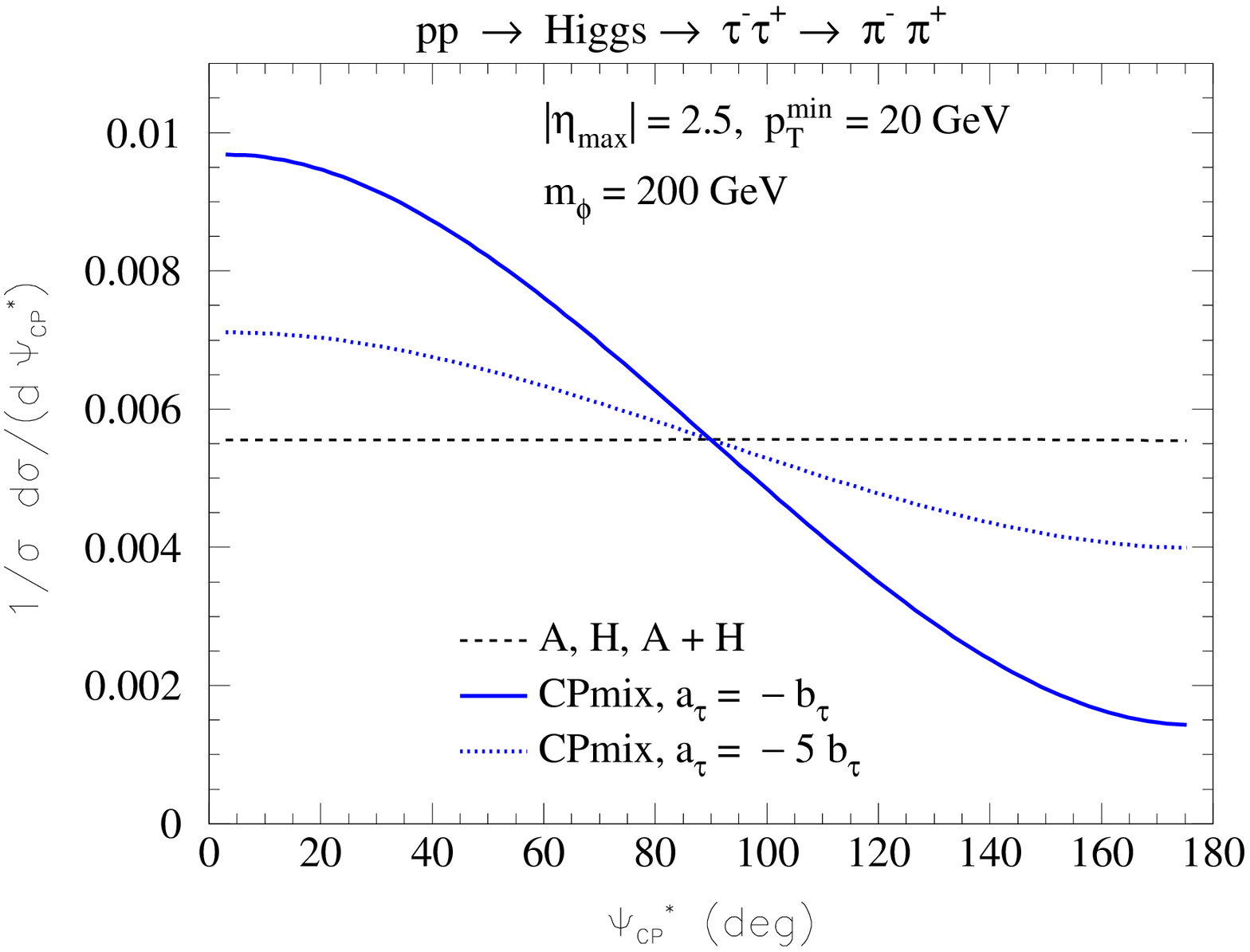}\vspace{-3pt}
 \par
\end{centering}
\caption{Distributions of $\varphi^{*}$ (a) and of $\psi_{CP}^{*}$ (b) for
several scenarios described in the text. In all cases the Higgs-boson
mass is taken to be $m_{\phi}=200$~GeV. \label{fig:phi_Ocp_mh200}}
\end{figure}

According to SM extensions, it is not unlikely that some of the Higgs-boson
states are (nearly) mass-degenerate. These states cannot be resolved
in the $\tau-$pair invariant mass spectrum. Suppose there is a scalar
and a pseudoscalar Higgs boson $H$ and $A,$ respectively, with nearly
degenerate masses which both contribute to the reaction (\ref{LHCphipi}).
The resulting distribution of the angle $\varphi^{*}$ (or of the
true decay-plane angle $\varphi_{true}^{*}$) will have a shape somewhere
between the scalar and pseudoscalar extremes shown in Figs. \ref{fig:2ab},
\ref{fig:3ab}, depending on the relative reaction rates. If such
a distribution would be found in an experiment one could, however,
not infer its origin. It could also be due to the production of one
(or several) $CP$ mixture(s) $\phi$ with mass(es) $m_{\phi}\approx m_{H,A}$.
This is shown in Fig. \ref{fig:phi_Ocp_mh200} (a), where the $\varphi^{*}$
distribution is plotted for two scenarios%
\footnote{For simplicity we consider in the following only one $CP$ mixture
$\phi.$ The case of several mass-degenerate Higgs boson states with
$CP$-violating couplings does not change our conclusions. %
}. Case (i): Production and decay of both a scalar and a pseudoscalar
Higgs boson with couplings such that the respective reaction rates
for (\ref{LHCphipi}) are equal, $\sigma_{H}=$$\sigma_{A}$.
This leads to a flat $\varphi^{*}$distribution. (Recall that interferences
of the $H$ and $A$ scattering amplitudes neither contribute to this
distribution nor to $\sigma.$) A distribution with the same shape
is, however, generated by a $CP$ mixture with scalar and pseudoscalar
couplings to $\tau$ leptons of equal strength, $|a_{\tau}|=|b_{\tau}|$.
This can be understood with the unsigned correlation (\ref{eq:dec1cor}).
The angle $\varphi_{true}^{*}$ is equally distributed for these couplings.
Case (ii): Here the $\varphi^{*}$ distribution is shown for $H$ and
$A$ exchanges with couplings such that $\sigma_{H}=2\sigma_{A}.$
This scalar-like distribution has the same shape as the one that originates
from the decay of a $CP$ mixture with couplings 
 $|a_{\tau}|=\sqrt{2}|b_{\tau}|$.
(Again, inserting these couplings into (\ref{eq:dec1cor}) results
in a $\varphi_{true}^{*}$ distribution which is practically identical
to the one shown in Fig. \ref{fig:phi_Ocp_mh200} (a).) For comparison
we have also plotted in Fig. \ref{fig:phi_Ocp_mh200} (a) again the
distributions due to a pure scalar and a pure pseudoscalar boson.

The two cases, $H+A$ exchange versus exchange of a $CP$ mixture,
can be disentangled with the $CP$ angle $\psi_{CP}^{*}.$ In Fig.
\ref{fig:phi_Ocp_mh200} (b) the distribution of this angle is shown
for $CP$-invariant and $CP$-violating Higgs-boson couplings. If
$CP$ is conserved, as it is the case for $H,$ $A,$ or $H+A$ exchange,
the expectation value of the $CP$-odd triple correlation associated
with $\psi_{CP}^{*}$ is zero. That is, the distribution of this observable
and, likewise, that of $\psi_{CP}^{*}$ is symmetric, the latter one
with respect to $\psi_{CP}^{*}=90^{\circ}.$ In fact, as Fig. \ref{fig:phi_Ocp_mh200}
(b) shows, $H,$ $A,$ or $H+A$ exchange leads to an essentially
flat distribution. On the other hand, for a $CP$ mixture the distribution
of the $CP$ angle is asymmetric with respect to $\psi_{CP}^{*}=90^{\circ}.$
Fig. \ref{fig:phi_Ocp_mh200} (b) shows the case of an ideal mixture
with couplings $a_{\tau}=-b_{\tau}$ and the case of a $CP$ mixture
where $b_{\tau}=-5a_{\tau}$. Notice that the case $a_{\tau}=-5$$b_{\tau}$
yields the same $\psi_{CP}^{*}$ distribution. This scalar-like Higgs
boson can be distinguished from the pseudoscalar-like boson $(b_{\tau}=-5a_{\tau})$
by means of the $\varphi^{*}$distribution. In addition to the $\psi_{CP}^{*}$
distribution, one may use the asymmetry
\vspace{-8pt}
\begin{eqnarray}
A_{CP} & = &
\frac{N(\psi_{CP}^{*}>90^{\circ})-N(\psi_{CP}^{*}<90^{\circ})}{N_{>}+N_{<}}
 \label{eq:A_CP}
\end{eqnarray}
\vspace{4pt}
in order to discriminate between $CP$-conserving and $CP$-violating
Higgs-boson exchanges.

How robust are the distributions of $\varphi^{*}$ and $\psi_{CP}^{*}$
with respect to measurement uncertainties expected at the LHC? In
order to study this question with Monte Carlo methods, we have accounted
for the expected measurement uncertainties by ``smearing'' the relevant
quantities with a Gaussian according to $\exp(-\frac{1}{2}(X/{\sigma)}^{2})$,
where $X$ denotes the generated quantity (coordinate in position
space, momentum component, energy) and $\sigma$ its expected standard
deviation (s.d.).

To obtain a rough idea about the length scales involved in the measurement
one may assume for a moment that the Higgs boson is produced at rest
in the laboratory frame. Then the energy of each $\tau$ lepton is
$m_{\phi}/2$. For a $2$-body decay of a $\tau$ lepton into a $\pi$
and a neutrino, the energy of the $\pi$ in the $\tau$-rest frame
is $E_{\pi}=m_{\tau}/2$. 
If one assumes that the pion is emitted transversely to the $\tau$  
direction, then the angle between the $\tau$  and the $\pi$  in the 
laboratory frame is $\angle_{lab}({\bf k},{\bf p)}\approx 29$~mrad, 
$17$~mrad, and $7$~mrad for $m_{\phi}$ = 120 GeV, 200 GeV, and 500 GeV, 
respectively. If one assumes that the decay length of a 
certain $\tau\to\pi\nu$  event is given by the average $\tau$  decay 
length, $c\tau_{\tau}=87\mu$m, the length of the impact-parameter
vector ${\bf n}$ in the laboratory frame is 
$|{\bf n}|\approx 80\,\,\mu{\rm m}$ for the three Higgs-boson masses. 
In view of this fact and in view of the relatively large value of 
$|{\bf n}|$  our method works
for a large range of Higgs masses. This rough estimate also indicates
the resolution that must be achieved in an experiment for the primary
vertex and the tracks of the pions.

The length of ${\bf n}$ depends on the decay length of the
$\tau$ lepton. For decay lengths shorter than the one used above,
$|{\bf n}|$ will be smaller and therefore smearing will affect
the distributions $\varphi^{*}$ and $\Psi_{CP}^{*}$ in a stronger
fashion. Our proposed distributions loose their discriminating power
for $\tau$-decay events with very short decay lengths. Using the
exponential decay law of the $\tau$ leptons in their rest frame,
we found in our numerical simulations that, for instance in the case
of $m_{\phi}=200$~GeV, a minimum decay length of $l_{\tau}^{min}=2$~mm
is required for both $\tau$ leptons in order to obtain reasonable
results. In an experiment such a cut could be realized by applying
a minimum cut on $|{\bf n}|$. Due to this requirement the number
of $\tau^{-}\tau^{+}$ events decreases approximately by a factor
of $2$. On the other hand, such a cut might be experimentally advantageous
to separate the $\tau\tau$ events from the background. 

The $\phi\to\tau^{-}\to\pi^{-}$ decay plane is illustrated in Fig.~\ref{fig:decplL},
and the $\phi\to\tau^{+}\to\pi^{+}$ plane may be drawn analogously.
In order to simulate the uncertainties in the experimental determination
of the production/decay vertex $PV$ and of the $\pi^{\mp}$ tracks,
we vary the position of $PV$ along and transverse to the beam axis
with $\sigma_{z}^{PV}=30\,\mu{\rm m}$ and $\sigma_{tr}^{PV}=10\,\mu{\rm m}$,
respectively. The track of a charged pion is smeared at the intersection
point of the impact-parameter vector ${\bf n}$ and the pion momentum
${\bf p}$. We vary this point within a circle of radius $\sigma_{tr}^{\pi}=10\,\mu{\rm m}$
transverse to the $\pi$ track. The angular resolution of the $\pi$
track is smeared by $\sigma_{\theta}^{\pi}=1\,{\rm mrad}$ around
the track. Moreover, the energy of the $\pi$ is varied by $\Delta E^{\pi}/E^{\pi}=5\%$.
These values appear to be realistic for the LHC experiments \cite{Gennai:2006,Tarrade:2007zz}.
As mentioned before, we apply also a minimum cut on the $\tau$~decay
length, which is $l_{\tau}\geq$$2\,{\rm mm}$ for $m_{\phi}=200$
GeV.

\begin{figure}[h]
\begin{centering}\includegraphics[clip,scale=0.42]{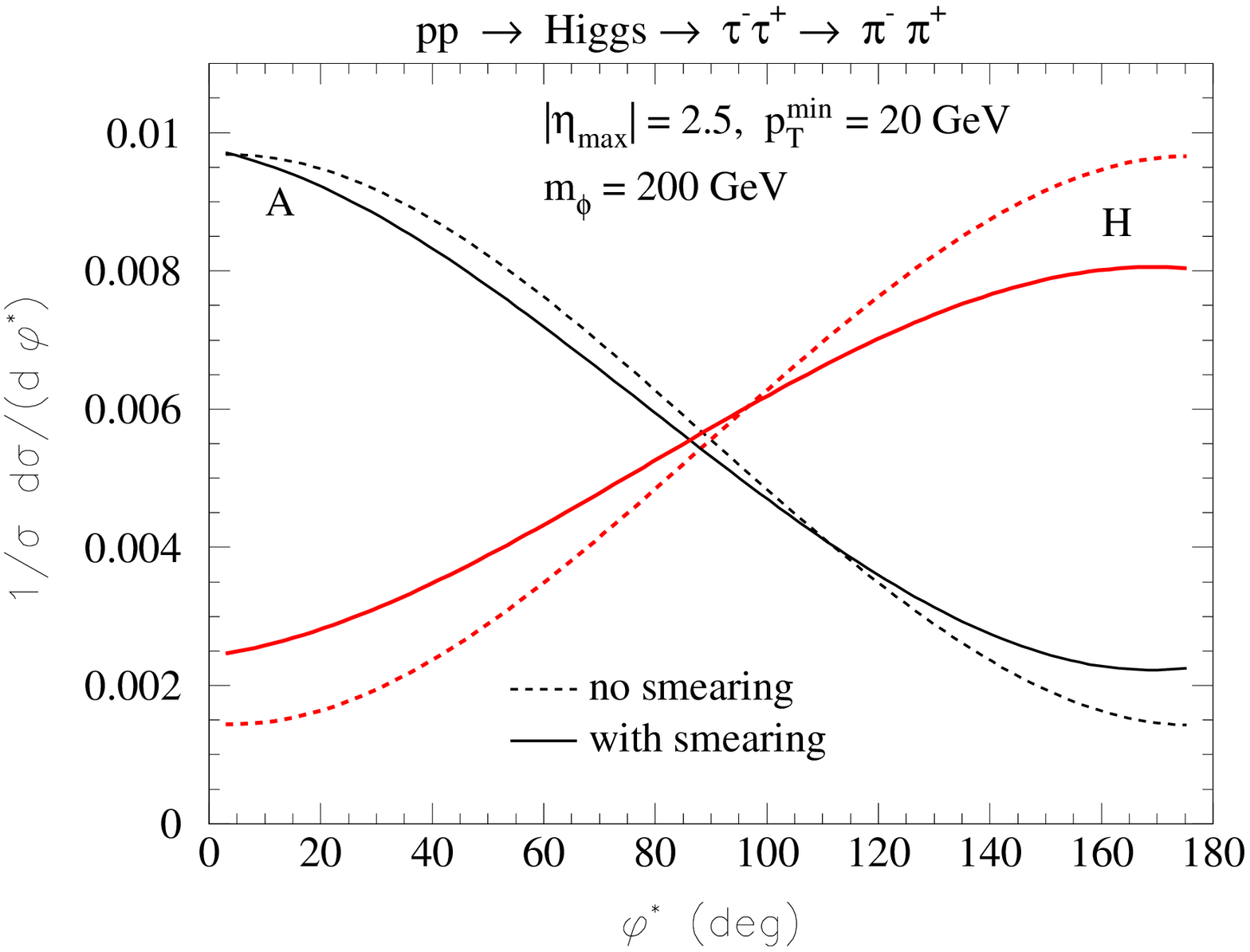}\hspace*{.4cm}%
\includegraphics[clip,scale=0.42]{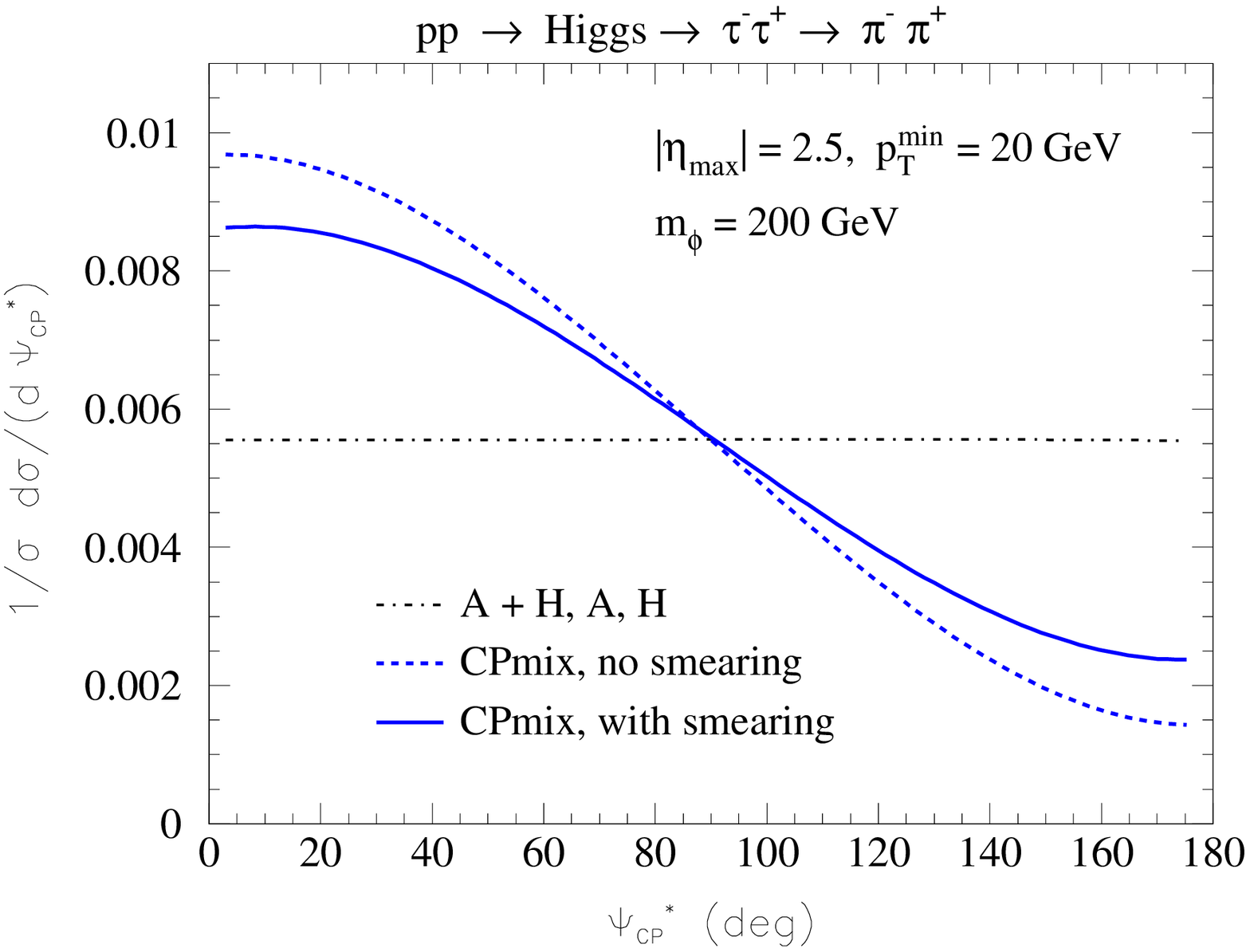}\vspace{-3pt}
 \par
\end{centering}
\caption{Distributions (a) for $\varphi^{*}$ and (b) for $\psi_{CP}^{*}$,
taking into account measurement uncertainties. In both figures the
dashed lines show the distribution without smearing, while the solid
lines include all smearing parameters as stated in the text. \label{fig:Smeared_mh200}}
\end{figure}

Taking this smearing into account, the distribution of the angle $\varphi^{*}$
is displayed in Fig.~\ref{fig:Smeared_mh200} (a) for a scalar and
a pseudoscalar Higgs boson with mass $m_{\phi}=200$ GeV.  
For comparison, this figure contains also the unsmeared distributions
already shown in Fig.~\ref{fig:2ab}. The solid lines show the distributions
using the smearing parameters stated above. While the simulated uncertainties
diminish the difference between the curves for $A$ and $H$ somewhat,
they are still clearly separated. This holds true also for other Higgs-boson
masses, as we have checked for $120$ GeV $\leq m_{\phi}\leq500$
GeV. Thus we conclude that $\varphi^{*}$is an appropriate observable
to distinguish between a scalar and a pseudoscalar Higgs boson at
the LHC.

Fig.~\ref{fig:Smeared_mh200} (a) shows that smearing affects the
$H$ distribution stronger than the distribution for $A$. If the
smearing parameters are chosen to be very large, the distributions
will peak at $\varphi^{*}\to0^{\circ}$ and will be depleted for large
$\varphi^{*}$; i. e., the distribution for $H$ will approach the
one for $A$. This is because for large values of $|{\bf n}_{-}|$
and $|{\bf n}_{+}|$ the minimum distance between the $\pi^{-}$ and
the $\pi^{+}$ tracks becomes negligibly small compared to $|{\bf n}_{\mp}|$
and therefore ${\bf n}_{-}$ and ${\bf n}_{+}$ will be almost parallel
in the $\pi\pi$ ZMF. 

The distribution of the $CP$ angle $\psi_{CP}^{*}$ is displayed
in Fig.~\ref{fig:Smeared_mh200}~(b). In the case of production
and decay of one or several (mass-degenerate) $CP$ eigenstates ($H$,
$A,$ $A+H,$ etc.), the distribution of $\psi_{CP}^{*}$, which is
a horizontal line, is not affected by any smearing. The dotted line
is the unsmeared distribution due to the decay of an ideal $CP$ mixture
($a_{\tau}=-b_{\tau}$), already displayed in Fig. \ref{fig:phi_Ocp_mh200}
(b). The solid line includes the uncertainties stated above. These
uncertainties decrease the discriminating power of $\psi_{CP}^{*}$
somewhat, but this observable is clearly the appropriate tool for
for distinguishing between Higgs-boson states with $CP$-conserving
and $CP$-violating couplings to $\tau$ leptons.

The distribution of the variables $\varphi^{*}$ and of $\psi_{CP}^{*}$
can be determined in completely analogous fashion also for the other
1-prong $\tau$ decays, that is, $\phi\to\tau^{-}\tau^{+}\to f_{1}^{-}\ f_{2}^{+}\:+\, neutrals$,
where $f_{1}^{-},$$f_{2}^{+}$ denote either a charged lepton from
$\tau^{\mp}\to e^{\mp},$$\mu^{\mp},$ or a charged pion from $\tau^{\mp}\to\rho^{\mp}$
and $\tau^{\mp}\to\pi^{\mp}2\pi^{0}.$ For $\phi=$$H,\, A$ the distribution
of the true decay-plane angle $\varphi_{true}^{*}$ in the $f_{1}f_{2}$
ZMF is given by $\sigma^{-1}d\sigma/d\varphi_{true}^{*}=(\pi)^{-1}[1\mp(\pi^{2}/16)\kappa_{1}\kappa_{2}\cos\varphi_{true}^{*}],$
where the numbers $\kappa_{1,2}$ signify the $\tau$-spin analyzer
quality of the respective charged prong. We recall that for the decays
of $100\%$ polarized $\tau$ leptons we have the angular distribution
$d\Gamma(\tau^{\mp}\to f^{\mp})$$\propto1$$\pm\kappa_{f}\cos\theta_{f}$,
where $\theta_{f}$ is the angle between the $\tau$ spin vector and
the direction of the charged prong in the $\tau$ rest frame. While
the spin analyzer quality factor is maximal, $\kappa_{\pi}=1,$ for
$\tau\to\pi\nu_{\tau}$, it is considerably smaller for the other
1-prong decays. For $\tau\to l=e,\mu$ it is $\kappa_{l}=-0.33$,
and for a charged pion from $\tau\to\rho$ it is only $\kappa\simeq-0.07.$
The spin analyzer quality of a charged lepton and of the charged pion
from $\rho,$ $a_{1}$ can, however, be significantly enhanced by
appropriate energy cuts, i.e., by taking into account only charged
prongs above some suitably chosen minimum energy in the $f_{1}f_{2}$
ZMF. The gain in spin-analyzing power outmatches by far the loss in
statistics. In this way, one can achieve an effective correlation
coefficient $\kappa_{eff}\simeq0.8$ while reducing the number of
1-prong $\tau$ decays that can be used in the analysis from $N_{1}$
to $N_{eff}\simeq0.54$$N_{1}$. Likewise, the discriminating power
of the $CP$-odd spin correlation underlying the distribution of $\psi_{CP}^{*}$
is maintained by these cuts. A detailed account will be given elsewhere
\cite{Berge:2008z}.

Finally, we make a crude estimate of how many events are needed in
order to distinguish between i) a scalar and pseudoscalar Higgs boson
and/or ii) between $CP$-conserving and $CP$-violating states, assuming
$m_{\phi}=200$ GeV. As to i), we define an asymmetry \begin{eqnarray}
A_{\varphi^{*}} & = & \frac{N(\varphi^{*}>90^{\circ})-N(\varphi^{*}<90^{\circ})}{N_{>}+N_{<}}\quad.\label{eq:A_phi}\end{eqnarray}

From Fig. \ref{fig:Smeared_mh200} (a) we obtain from the smeared
distributions $A_{\varphi^{*}}^{H}=0.31$ and $A_{\varphi^{*}}^{A}=-0.42$.
Taking into account an effective $\tau$-spin analyzing coefficient
for 1-prong decays $\kappa_{eff}=0.8,$ these 
 asymmetries are reduced by a factor
of about $0.64.$ Thus, for distinguishing $H$ from $A$ with 5 s.d.
significance requires about 120 1-prong events. Concerning ii), the
result of Fig. \ref{fig:Smeared_mh200} (b) implies that for an ideal
$CP$ mixture the $CP$ asymmetry defined in Eq.~(\ref{eq:A_CP})
takes the value $A_{CP}=-0.37$ while it is zero for pure $H,$ $A$
and degenerate $H$ and $A$ intermediate states. Thus, about 400
1-prong events will be needed to establish this $CP$-violating effect
at the 5 s.d. level. This should be feasible, depending on the masses
and couplings of $\phi$, after several years of high luminosity runs
at the LHC.

\section{Conclusions}

We have proposed a method for determining the $CP$ nature of a neutral
Higgs boson or spin-zero resonance $\phi$ at the LHC in its $\tau$
pair decay channel. The method can be applied to any 1-prong decay
mode of the $\tau$ lepton. It requires the measurement of the energy
and momentum of the charged prong ($\pi^{\mp},$ $e^{\mp},$ $\mu^{\mp})$
from $\tau^{\mp}$ decay and the determination of the $\tau^{-}\tau^{+}$
production vertex with some precision. The distributions of the angles
$\varphi^{*}$ and $\psi_{CP}^{*}$ allow to discriminate between
pure scalar and pseudoscalar states and/or between a $CP$-conserving
and $CP$-violating Higgs sector. For the decays $\tau\to\pi\nu_{\tau}$
we have shown that the variables $\varphi^{*}$ and $\psi_{CP}^{*}$
maintain their discriminating power when measurement uncertainties
are taken into account. The smearing parameters that we used in our
simulations indicate the precision which should eventually be achieved
 by the LHC experiments. Our method could, of course, be applied also
at a future $e^{+}e^{-}$ linear collider where Higgs-boson production
and decay would take place in a much cleaner environment.

\section*{Acknowledgements}

We thank P. Sauerland and A. Stahl for helpful discussions. The work
of \,S.\,B. is supported by the Initiative and Networking Fund of
the Helmholtz Association, contract HA-101 (`Physics at the Terascale')
and the work of \,W.\,B. is supported by Deutsche Forschungsgemeinschaft
SFB/TR9.

\end{document}